# $^{133}$Cs$^{87}$Rb LASER COOLED DUAL FOUNTAIN

D. Calonico[1], Y. Sortais[2], S. Bize[2], H. Marion[2], C. Mandache[3], C. Salomon[4], G. Santarelli[2], A. Clairon[2]

[1]Politecnico, Corso Duca degli Abruzzi 24, 10100 Torino, Italia
[2]BNM-SYRTE, 61 Av. de l'Observatoire, 75014 Paris, France
[3]Institutul National de Fizica Laserilor, Plasmei si Radiatiei, P.O. Box MG36, Bucaresti, Magurele, Romania
[4]Laboratoire Kastler Brossel, Ecole Normale Supérieure, 24 rue Lhomond, F-75231 Paris, France
Corresponding author : calonico@tf.ien.it

## 1. ABSTRACT

At SYRTE in Paris two Cs fountains and a $^{87}$Rb one have been already demonstrated. [Refs 1, 2]. This paper illustrates the construction of a $^{87}$Rb and Cs dual fountain realized within the same structure to work simultaneously. Up to now only the Cs part is under operation but Rb part, already completed, will be soon operational. The experiment has been projected to improve present fountain standards and to perform a more precise laboratory test, at $10^{-16}$ level, on a possible time variation of fine structure constant $\alpha$ as recent cosmological evidences strongly motivates [Ref. 3].

## 2. DUAL FOUNTAIN AND TIME VARIATION OF THE FINE STRUCTURE CONSTANT

In Prestage theoretical framework [Ref. 4] experiments have been already performed to measure an eventual time variation of fine structure constant $\alpha$. Previous measurements using Cs and Rb LPTF fountains have shown no $\alpha$ variation at $7 \times 10^{-15}$ per year level. Dual fountain project had been developed to take advantages of the same experimental structure to reach a more accurate evaluation. As C-field and black-body temperature are the same, their fluctuations are strongly correlated. Their contribution to the uncertainty of the comparison is then rejected at 56% and 79% level respectively [Ref. 9]. Finally, advantages will come from a simpler synchronisation of the fountains sequences which will result in a better rejection of the interrogation oscillator noise and synchronous perturbations. These are the main reasons that make possible an evaluation of $\alpha$ variation at the $10^{-16}$/year level.

## 3. EXPERIMENTAL APPARATUS

The dual fountain has a (1,1,1) configuration. Two optical benches, one for Rb and one for Cs, provide all the radiations required. They are completely separated and independent each other and from the fountain structure. The benches are coupled to the vacuum chamber by polarizing fibres through prealigned collimators. Dichroic plates will be used to mix 780 and 852 nm laser beams at the input of the collimators.
In the same trapping region Rb and Cs molasses are loaded by two atomic beams, pre-cooled by a chirping laser technique. Also the chirped laser benches are independent and coupled by fibres to the structure.
Cs optical bench is composed of a master and a repumper both SDL cavity extended diode laser locked by using saturated absorption technique. Optical power amplification is done by injection locking of two slave SDL diodes by the master.
Detection radiation is provided by the repumper itself and by an indipendent low noise SDL extended cavity diode laser locked by using an FM spectroscopy technique and a fast servo system. The detection power is servo controlled by an acousto-optic modulator. So far, contribution to the frequency instability from laser sources is $\sigma_y(\tau) = 9 \times 10^{-16} \tau^{-1/2}$ [Ref. 8].
We collect up to $10^9$ Cs atoms and detect up to $10^7$ atoms by 0.3 s loading time. Atomic beam flux is about $10^{12}$ atoms/s at temperature of the oven of 383 K.
Ramsey fringes are shown in figure 1: launching height is typically 0.865 m above trapping region, with a corresponding 0.53 s between the two microwave interactions producing 0.94 Hz FWHM Ramsey fringes. A really good S/N ratio of 3000 was measured with $\pi/4$ pulses technique.
The Rb optical bench is quite similar to the Cs one, except for the slave laser, which is a MOPA amplifier.

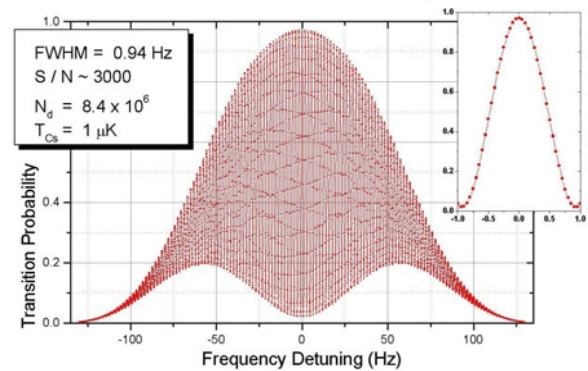

Figure 1. Cs Ramsey Fringes on clock transition; atoms detected about $10^7$, S/n = 3000, FWHM =0.94 Hz (Ramsey time 0.53 s).

Two systems insure temperature control of the microwave interaction region. First, an aluminium envelope wounded with twisted Arcap heating wire, separated by a μ-metal shield from the drift tube can work continuously without perturbing atoms. It maintains the temperature at about 299 K. The second system ensures the fine cavity tuning temperature of 299.12 K. It consists of Arcap twisted wire wounded directly on vacuum drift tube to minimize the time constant of thermal exchanges. So far, this system is pulsed, switched on only during trapping time by a 100 kHz AC to avoid any perturbation for the atoms.
Temperature is monitored by three 100 ohms platinum sensors at different heights of the tube and a servo loop is going to be put under operation to perform a 0.1 K control. Up to now, a 1-2 K fluctuation is insured, leading to an accuracy contribution of the black-body radiation shift of 2.5-$5\times10^{-16}$.

Magnetic field homogeneity and control is provided by four compensation coils, orthogonal to the tube and distributed along its height. In usual operation we generate a field of about 200 nT with 0.1 nT inhomogeneity as shown in figure 3. This homogeneity gives a contribution of $3\times10^{-16}$ to accuracy budget.

Another control consists of four coils also distributed along the tube height which provide a field of 400 nT in the cooling region. This field is servo controlled. The magnetic sensor is housed at the height of the trapping region. Extra coils are installed along the atoms path in order to avoid non adiabatic spatial variations of the magnetic field

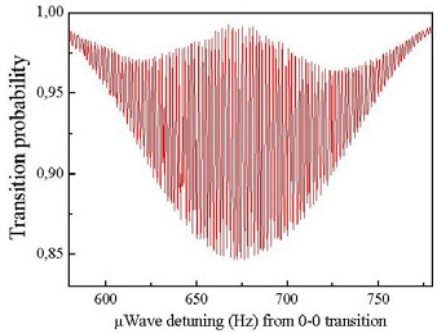

Figure 2. Ramsey Fringes on magnetic $m_F=1$-$m_F=1$ transition, FWHM =0.94 Hz

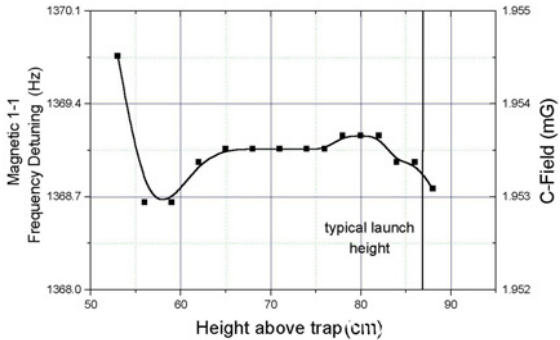

Figure 3. C-field map, showing homogeneity at 0.1 nT level

Cs and Rb interrogation cavities are realized from the same monolithic copper block to have exactly the same temperature and then reject thermal fluctuation in comparison between Rb and Cs hyperfine frequency. They sustain the $TE_{011}$ mode and they are both symmetrically fed through two iris by lateral rectangular cavities. These latter are both coupled to coaxial cable by an antenna. The $TE_{011}$ loaded Q factor is about 7000 in both cavities, while the degenerated $TM_{111}$ is strongly coupled by a piece of coaxial guide (anti resonant with $TE_{011}$) that prolongs each cavity, to detune its resonant frequency of about 100 MHz and to reduce its Q factor less than 1000. To minimize microwave leakage a brazing and indium o-rings are provided.

The tuning of the cavities is done by fine polishing. We have measured that at the temperature of 299.12 K the two cavities are resonant at their respective atomic frequencies with a detuning of less than than 10 kHz. This is quite important in order to minimize the cavity pulling effect, that a dual fountain can not reject.

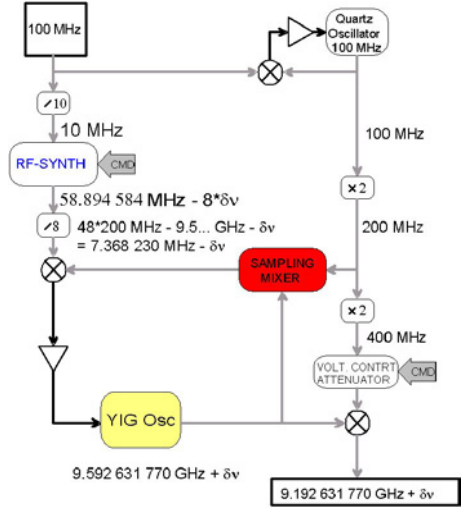

Figure 4. Synthesis chain for Cs hyperfine frequency.

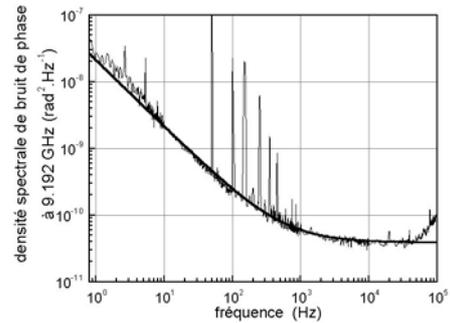

Figure 5. Phase noise power density of Cs synthesis chain.

The synthesis chain scheme is showed in figure 4. Its phase noise spectral density at 9.192 GHz (figure 5) is described between 1 Hz and 30 kHz by the relation

$$S_\phi(f) = 2.1\times10^{-8}f^{-1} + 3.9\times10^{-11} \text{ rad}^2\text{Hz}^{-1}$$

That leads to a contribution to stability in term of Allan deviation of $\sigma_y(\tau) = 2.5\times10^{-14}\tau^{-1/2}$.

As local oscillator we have used either a BVA quartz or a Sapphire Cryogenic Oscillator (SCO) both locked on H-Maser. The SCO has been realized at University of Western Australia [Ref. 6,7]. Now its performances has been improved by some changes in its cryogenic structure and power servo-loop. A stability of $3\times10^{-15}$ up to 1000 s is now available, then a thermal derive starts of about $10^{-13}$/day. In figure 7 we show the Allan deviation measured using the SCO free running as local oscillator of the fountain operating in Cs mode only, before and after improvements. An unprecedent $\sigma_y(\tau) = 3.6\times10^{-14}\tau^{-1/2}$ up to 20 s was obtained, while after changes we had measured a $\sigma_y(\tau) = 5\times10^{-14}\tau^{-1/2}$ up to 1000 s. The worst result at 1 s is due to an excess noise of the fountain during the second run of measurements. We are investigating about it. Up to know this stability seems to be limited by the fountain itself. In figure 6 is shown the Allan deviations measured with BVA and SCO locked on H-Maser as local oscillators. In the first case the stability of the fountain is limited by the BVA at $\sigma_y(\tau) = 1.1 \times 10^{-13} \tau^{-1/2}$. In the second case the H-Maser limits the fountain stability in short and medium term at $\sigma_y(\tau) = 6.6 \times 10^{-14} \tau^{-1/2}$.

All of this measures were obtained detecting about $10^7$ Cs atoms using a time-cycle of 1.3 s and a FWHM for the central Ramsey fringe of 0.94 Hz.

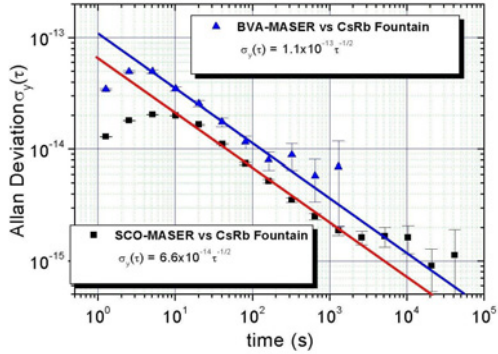

Figure 6. Dual fountain stability (Cs operated), with two different local oscillator. Squares: SCO locked on H-Maser, Allan deviation $\sigma_y(\tau) = 6.6 \times 10^{-14} \tau^{-1/2}$; Triangles: BVA locked on H-Maser, $\sigma_y(\tau) = 1.1 \times 10^{-13} \tau^{-1/2}$

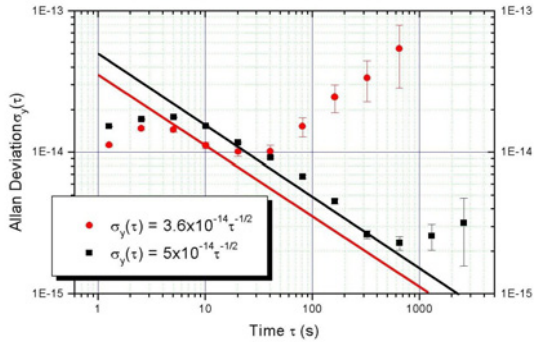

Figure 7. Dual fountain stability (Cs operated), SCO free running as local oscillator. Circles: SCO before improving, Allan deviation $\sigma_y(\tau) = 3.6 \times 10^{-14} \tau^{-1/2}$ up to 20 s; Squares: SCO after improving, $\sigma_y(\tau) = 5 \times 10^{-14} \tau^{-1/2}$ up to 1000 s. See text for $\sigma_y(1)$ deterioration.

## 4. CONCLUSION AND PERSPECTIVES

In figure 8 it is shown the preliminary accuracy budget of dual fountain when operated Cs. The preliminary evaluation gives an accuracy of $1.2 \times 10^{-15}$, a better evaluation of the accuracy will be performed soon. In these measurements the fountain was run with about $10^6$ atoms detected and Allan deviation of $\sigma_y(\tau) = 1.4 \times 10^{-13} \tau^{-1/2}$.

Collisional shift uncertainty has been evaluated at 30% level: this is conservative and work is in progress to reach a stable 10% level. It seems to be hard but not impossible to get a few percent level in order to attend $10^{-16}$ accuracy region. Also black body shift uncertainty is conservative waiting for implementation of servo loop to get 0.1 K fluctuation. This will improve to about $3 \times 10^{-17}$ contribution to accuracy.

First comparison with Pharao Cs fountain at SYRTE shows good agreement at $1\sigma$ level ($\sigma = 1.5 \times 10^{-15}$) but more measures have to be taken.

So far, the preliminary evaluation for Cs shows no particular problems on fountain structure, in magnetic field, thermal gradient, microwave part. If we consider the same situation, for Rb we obtain an expected accuracy of $6 \times 10^{-16}$.

$^{87}$Rb part of the dual fountain is now in progress: the apparatus is almost completed and we are going to put it under operation. A straightforward precise evaluation of accuracy both for Cs and Rb will be performed, together with a new measure of the two hyperfine frequencies ratio. From the Cs preliminary evaluation we can expect for Rb a better accuracy than in the previous Rb fountain where the $10^{-15}$ level was reached and in which the strongest limitation was due to microwave leakage [Ref 2].

| Effect | Experimental testing procedure | C [$10^{-15}$] | UB [$10^{-15}$] |
|---|---|---|---|
| 2nd order Zeeman | Field map | -178 | 0.2 |
| Blackbody Radiation | DC Stark + direct measurement | 17.4 | 0.4 |
| Microwave spectrum | μWave power and Timing (Differential Measurement) | 0 | 0.2 |
| Microwave leakages | μWave power (Diff. Meas.) | 0 | 0.2 |
| Pulling by other lines | μWave power C-field (Diff. Meas.) | 0 | 0.1 |
| Gravitational Red Shift | ---- | 6.5 | <<0.1 |
| Cold collisions | Density (Diff. Meas.) | 4 | 1 |
| | | Total UB | $1.2 \times 10^{-15}$ |

Figure 8. Dual Fountain Cs operated preliminary evaluation.